\documentclass[]{article}
\newcommand{\bc}{\begin{center}}
\newcommand{\ec}{\end{center}}
\newcommand{\half}{\frac{1}{2}}
\newcommand{\be}{\begin{equation}}
\newcommand{\ee}{\end{equation}}
\newcommand{\bea}{\begin{eqnarray}}
\newcommand{\eea}{\end{eqnarray}}
\newcommand{\nn}{\nonumber}

\begin{document}


\title{{\small\begin{flushright}
ITFA--2000--04 \\  
\end{flushright}} Effect of CP-violation on the sphaleron rate}

\author{Bert-Jan Nauta\thanks{mail: 
nauta@phys.uva.nl}\\
Institute for Theoretical Physics, University of Amsterdam\\
        Valckenierstraat 65, 1018 XE  Amsterdam, The Netherlands}

\maketitle

\begin{abstract}
We calculate the effect of two CP-violating (dimension-eight) operators 
of the SU($2$)-Higgs model on the 
motion along a particular path from the vacuum to the sphaleron. 
It turns out that
CP-violation may introduce a difference between the sphaleron rate towards 
larger Chern-Simons number and the rate towards smaller Chern-Simons number. 
Such a difference induces a non-zero baryon-number without 
a first order phase transition.
\end{abstract} 
The matter anti-matter asymmetry of the present universe is
an important cosmological observation. 
It can be given a quantitative meaning by considering the ratio of the net
baryon-number density and photon density \cite{observation}
\be
\frac{\Delta_{B}}{n_{\gamma}}\sim 10^{-10},
\label{observationratio}\ee
with \(\Delta_{B}=n_{B}-n_{\bar{B}}\)
the difference between the baryon-number density and the anti-baryon-number 
density.
This ratio is constant under the expansion of the universe. Contrary to its 
superficial appearance the  baryon-number excess (\ref{observationratio})  is 
actually very large to be explained by the standard model.
The problem is to explain the generation of this amount of baryons, when the 
universe started out from a state with baryon-number equal to zero
(see for recent reviews e.g. 
\cite{rubakov,riotto}). 
In 1967 Sakharov \cite{sakharov} was the first to address this problem and he 
noted that there are three prerequisites, namely\\
1) non-conservation of baryon-number, \\
2) C- and CP-violation,\\
3) departure from equilibrium.\\
In relation to the first requirement,
it was discovered by 't Hooft \cite{hooft} that in the standard model 
the baryon 
number is not conserved,  as a consequence of  the anomaly equation
\be
\partial_{\mu}j_{B}^{\mu}=\frac{3g^2}{32\pi^2}F_{\mu\nu}^a \tilde{F}^{\mu\nu a},
\label{anomaly}\ee
and  
transitions between different vacua that are labeled 
by the Chern-Simons number. 
A transition from one (classical) vacuum to the next in the positive 
(negative) Chern-Simons direction
yields a change in baryon-number of \(+ 3 (-3)\).
These transitions are 
possible because of quantum tunneling through the barrier separating the 
different vacua, but at zero temperature they are very  much suppressed. 
At high temperatures the system can go over the 
barrier due to thermal fluctuations, because of that the rate of 
baryon-number 
changing transitions can be quite large. The transition rate is 
proportional to the Boltzmann factor:
\(\Gamma_{\rm sph}\sim \exp -E_{\rm sph}/T\) \cite{klinkhamer,kuzmin,arnold},
with the energy of the sphaleron 
\(E_{\rm sph}= \mbox{number}\times 4\pi v /g\),
where \(v\) is the expectation value of the Higgs field at temperature \(T\).
It is the sphaleron energy that occurs in the Boltzmann factor since 
the sphaleron is the minimal energy configuration at the barrier.

In the standard model CP-violation occurs in the CKM-matrix; however it is 
too small to explain the observed number of baryons. This is 
an indication for physics beyond the standard model with extra CP-violation 
(such as the two Higgs doublet model or the minimal supersymmetric standard 
model). 

Also it has been established that in the standard model there is no (strong)
 first-order electroweak phase transition \cite{kajantie}. 
This also has been taken as an 
indication for new physics that should provide a 
strong enough departure from equilibrium. 
The requirement for a strong first-order 
electroweak phase transition has been used to 
constrain 
parameters of extensions of the standard model \cite{cline}. Also new 
mechanisms for 
a departure from equilibrium at the electroweak scale have been considered,
see for example 
\cite{garcia-bellido1}.

In this letter we want to point out the possibility that in a model 
with sufficient CP-violation, the baryon-number expectation value
is non-zero when the system is in kinetic equilibrium, but
sectors with a different baryon-number are not in equilibrium.  
A non-zero expectation value of the B-number can occur
when the rate of sphaleron transitions to the vacuum
with a larger Chern-Simons number, \(\Gamma^{\uparrow}\), 
is different from the rate of transitions to the vacuum with a smaller
Chern-Simons number \(\Gamma^{\downarrow}\) (an example of such a difference 
out of equilibrium is discussed in \cite{garcia-bellido2}). 
If this is the case, an initial state 
with zero baryons will evolve into a state with non-zero baryon-number.

The question is how  CP-violating interactions may induce such a 
difference in rates. We discuss this for a specific CP-violating
action
\be
S_{CP}=\int d^4x \frac{1}{M^4}\left[ \delta_{CP}^{1}(D_{\rho}\phi)^{\dagger}
(D^{\rho}\phi)-\delta_{CP}^{2}\frac{1}{4}F_{\rho\sigma}^{a}F^{\rho\sigma a}
\right] \frac{3g^2}{32\pi^2}
F_{\mu\nu}^{b}\tilde{F}^{\mu\nu b}.
\label{CPviolaction}\ee
This action may be thought of to come from integrating out new (CP-violating) 
physics at the mass-scale \(M\) (which may be temperature-dependent, for 
instance through the temperature-dependence of \(v\)). 
The operators in (\ref{CPviolaction}) are the lowest dimensional CP-odd 
operators in the SU($2$)-Higgs sector that will contribute to the baryon 
number expectation value. We will see that the dimension-six operator 
\(\phi^{\dagger}\phi F\tilde{F}\) does not give a contribution.

To study the effect of CP-violation on sphaleron transitions, we consider the 
motion along a specific path starting at the vacuum and ending at the sphaleron
at \(N_{CS}=+ 1  /  2\) . For simplicity we use the path introduced in  
\cite{manton}. This path is not the minimal energy path, which 
was constructed in \cite{akiba}. But 
we expect that the precise path will not be 
important for the following rather general arguments and that the final 
result is sufficient as an order of magnitude estimate.
We parameterize 
the path by the (time-dependent) angle \(\Theta\in [0, \frac{1}{2}\pi]\), 
and use the following 
Ansatz for the fields (in the radial gauge)
\bea
A_{\mu}^{a}\sigma^{a}&=&
\frac{-2i}{g}f(r)[\partial_{\mu}U(\Theta)]U^{-1}(\Theta),
\label{gpathparametrization}\\
\phi&=&\half\sqrt{2} v h(r) U(\Theta)
\left(\begin{array}{c} 0 \\ 1\end{array}
\right), 
\label{hpathparametrization}\eea
with the \(\Theta\)-dependent SU($2$)-matrix 
\be
U(\Theta)=\frac{1}{r}\left(\begin{array}{cc} z & x+iy \\ -x+iy & z \end{array}
\right)\sin\Theta
+\left(\begin{array}{cc} i & 0 \\ 0 & -i \end{array}
\right)\cos\Theta
. 
\ee
This parameterization is a non-static generalization of the fields considered 
in
\cite{manton,klinkhamer} (with the identification \(\mu=\Theta\)). 
This 
particular generalization is convenient, since the field strength vanishes at 
infinity for the asymptotic boundary condition 
\(r\rightarrow \infty\) \(f(r)\rightarrow 1\).

We insert the fields (\ref{gpathparametrization}) and 
(\ref{hpathparametrization}) in the SU(2)-Higgs action
and the CP-violating action (\ref{CPviolaction}) and find
\bea
S&=&\int dt \left[\frac{ 4 \pi v^2}{(gv)^3}\left(a_1+a_2\sin^2 \Theta\right)
\dot{\Theta}^2
-\frac{4\pi v}{g}\left( a_3\sin^2\Theta + a_4 \sin^4\Theta\right)\right],
\label{actiontheta}\\
S_{CP}&=& \frac{4\pi v^2}{M^4}\int dt 
\left( b_1 \delta_{CP}^{1}+ b_2 \delta_{CP}^2 +b_3\delta^2_{CP}\sin^2\Theta\right) 
\dot{\Theta}^3 \sin^2\Theta.
\label{CPactionstheta}
\eea
In the CP-violating action we have neglected total time-derivatives. 
Had we included the dimension-six operator
\(\phi^{\dagger}\phi F\tilde{F}\) it would only have given a total 
time derivative.
The coefficients \( a_1\), \(a_2\), \(a_3\), \(a_4\), \(b_1\), \(b_2\), and 
\(b_3\) stand for integrals involving \(f(r)\), \(h(r)\),
\(\partial_{r} f(r)\), and  \(\partial_{r} h(r)\).
We use Ansatz b of \cite{klinkhamer} for the functions \(f(r)\) and \(h(r)\); 
then the parameters only 
depend on the ratio \(\lambda/g^2\), with \(\lambda\) the Higgs self-coupling.
We take \(\lambda=g^2\), for \(g\approx 0.65\) this sets the (zero temperature)
Higgs  mass at \(230\) GeV, and find for the coefficients the numerical values
\bea
a_1&=&2.51,\nn\\
a_2&=&1.35,\nn\\
a_3&=&1.58,\nn\\
a_4&=&0.53,\nn\\
b_1&=&0.14,\nn\\
b_2&=&0.096,\nn\\
b_3&=&0.23.
\label{coefficients}
\eea
The CP-odd action (\ref{CPactionstheta}) introduces a velocity-dependent 
force in the equations of motion. For the moment 
we ignore the \(\sin\Theta\)-dependence in (\ref{CPactionstheta}).
Then the force points in the direction of motion when the system moves from 
the vacuum towards the sphaleron at \(N_{CS}=+\frac{1}{2}\), whereas the 
force is opposite to the direction of motion when the motion is towards the 
sphaleron at \(N_{CS}=-\frac{1}{2}\). As a consequence, the system will
find it easier to cross the barrier to the right than to the left.
Therefore we expect that the probability of crossing the barrier to the right,
\(P^{\uparrow}\), is larger than the probability of crossing the barrier to 
the left, \(P^{\downarrow}\).

To obtain a quantitative estimate for the effect of the CP-odd terms
on the motion over the barrier, we consider the shift in the energy caused by
the extra CP-violating terms (\ref{CPactionstheta}) 
\be
E_{CP}(\Theta,\dot{\Theta})=\frac{8\pi v^2}{M^4}
\left( b_1 \delta_{CP}^{1}+ b_2 \delta_{CP}^2 +b_3\delta^2_{CP}\sin^2\Theta\right)
\dot{\Theta}^3 \sin^2\Theta.
\label{energyshift}\ee
Especially the typical energy shift at the sphaleron configuration
is important.
To calculate this energy shift,
we need the typical velocity \(\dot{\Theta}\). 
To zeroth-order in \(\delta_{CP}^1\) and \(\delta_{CP}^2\) 
the velocity is Gaussian distributed at the sphaleron and we find 
\be
\langle\dot{\Theta}^2\delta(\Theta-\frac{1}{2}\pi) \rangle=
\frac{(gv)^3 T}{4\pi v^2(a_1+a_2)},
\label{velocityest}\ee
where the \(\delta\)-function enforces that the average over the velocity
is taken at the sphaleron configuration.
With this estimate for the velocity we find for the typical 
energy shift
\be
\delta E_{\rm sph}=\frac{1}{\sqrt{\pi} v M^4}\left( b_1 \delta_{CP}^{1}+ 
b_2 \delta_{CP}^2 +b_3\delta_{CP}^2\right)
\left[\frac{(gv)^3T}{(a_1+a_2)}\right]^{\frac{3}{2}},
\label{energydif}\ee
which provides a quantitative measure for the amount of CP-violation.

As an  estimate for \(P^{\uparrow}\) we may take the probability that a 
configuration at the barrier moves in the positive Chern-Simons direction
\be
P^{\uparrow}=\langle\delta(\Theta -\frac{1}{2}\pi)H(\dot{\Theta})\rangle,
\ee
where \(H(\dot{\Theta})\) is the Heaviside function. In a similar manner 
\(P^{\downarrow}\) can be calculated. We get 
\be
P^{\uparrow\; (\downarrow)}=\frac{1}{2}+ (-) 0.80\; \beta \delta E_{\rm sph}.
\label{probs}\ee

We are interested in the case that \(T<<E_{\rm sph}\). Then the time that is 
spend  rolling down is much smaller than the time spend around the vacuum in 
between two barriers. In this case
 we may neglect the effect of noise (from the other 
degrees of freedom that were not taken into account in our 
\(\Theta\)-analysis) during the motion from one vacuum to the next.
And the difference in rates towards negative or positive Chern-Simons number is approximately the difference in the probabilities (\ref{probs})
\be
\Gamma^{\uparrow\;(\downarrow)}=\Gamma_{\rm sph}
\left[1+ (-)\; c \beta \delta E_{\rm sph}\right],
\label{CPrate}\ee
where $c$ is a coefficient of order one.
In the estimate for the upward and downward sphaleron 
rates (\ref{CPrate}) in the presence of CP-violating interactions 
(\ref{CPviolaction}) an uncertainty arises from 
the path that we have chosen, because the 
fields (\ref{gpathparametrization}) and (\ref{hpathparametrization}) 
do not satisfy the (SU($2$)-Higgs) 
equations of motion. However, for \(\Theta=\frac{1}{2}\pi, \dot{\Theta}=0\) 
these fields  
provide a very good approximation to the solution of the 
(static) field equations \cite{klinkhamer}.
Hence, we expect that
 close to the sphaleron and for small velocities \(\dot{\Theta}<< gv\),
the estimates (\ref{energydif}) and (\ref{CPrate}) provide a reasonable 
approximation.
In any case, the parametric dependence on 
\(g\), \(v\), \(M\), and \(T\) should be correct.

The difference in rates (\ref{CPrate})
has been obtained by treating a single transition as 
a classical motion over the barrier. The dynamics at a larger scale involving 
more transitions is different, namely that of a random walk 
with different probabilities of moving left or right.
This difference in rates or probabilities implies that the expectation value 
of the Chern-Simons number grows linearly in time
\be
\langle N_{\rm CS}(t)- N_{\rm CS}(t_{\rm in})  \rangle = V 
\left( \Gamma^{\uparrow}-\Gamma^{\downarrow}\right) (t-t_{\rm in}),
\label{lingrowth}
\ee
with $V$ the volume. The brackets denote a classical average over initial 
conditions with a normalizable probability distribution.
Note that (\ref{lingrowth})  is CPT-invariant. 

When we include the baryons into the system there is no infinite growth
of the baryon-number expectation value,
because there is an effective potential of the baryon-number that opposes 
the effect.
For small baryon-number densities the potential is quadratic: 
\(V_{\rm eff}(\Delta_{B})\sim \Delta_{B}^2\). Therefore also a non-zero 
baryon-number will induce a difference in rates \cite{arnold,khlebnikov}. 
Combining
the effect of a non-zero baryon-number density and CP-violation
to first order, we find for the rates
\be
\Gamma^{\uparrow (\downarrow)}(\Delta_B)=\Gamma_{\rm sph}
\left[1 - (+) \; 0.80 \frac{\Delta_{B}}{n_{\gamma}} + (-)\; 
c \frac{\delta E_{\rm sph}}{T}\right],
\label{barrate}
\ee
where \(n_{\gamma}=0.24 \; T^3\) is the photon density.
The rate equation is
\be
\frac{\mbox{d} \Delta_{B}}{\mbox{d}t}= 3 \left[\Gamma^{\uparrow}(\Delta_{B})-
\Gamma^{\downarrow}(\Delta_{B})\right].
\label{rateeq}\ee
{}From the rate equation  we find the stationary (and stable)
solution
\be
\frac{\Delta_B}{n_{\gamma}}= 1.25 c \frac{\delta E_{\rm sph}}{T}.
\label{barnumber}\ee

Also of interest is the width in the distribution of baryon-number densities. 
When the mean value and the width
are small compared to \(T^3\) 
the width of the distribution increases through diffusion.
Hence the width is of order \([\Gamma_{\rm sph} 
(t-t_{\rm in})/V]^{1/2}\) at time \(t\). We see that it is suppressed 
by the volume of the system. We can also determine the time-scale of 
equilibration of the system. 
We expect that the system starts to equilibrate when 
the distribution reaches the upperbound provided by
energy conservation of the classical subsystem.
The asymmetry (\ref{barnumber}) will then decrease 
and eventually vanish,
as it should in equilibrium (see e.g. \cite{riotto}).
The energy of the classical system is of order \(VT^4\).
We find that the time
where the width is of order \(T^3\) determines
the time-scale of equilibration. Using the estimate for the width given above,
we find that the equilibration 
time diverges in the infinite volume limit.

From the results obtained above a scenario for baryon-number generation in 
the universe may  be constructed.
Consider the situation that the universe at some time before the electroweak 
phase transition (or cross-over)
is in a state with baryon-number density equal to zero. This initial 
condition may be provided by inflation. 
Since the equilibration time is extremely long the broken phase will
then be entered with zero baryon-number density.
In the broken phase the value (\ref{barnumber}) 
will be reached in a relatively short time.
As the universe expands and the temperature decreases, the baryon-photon
 ratio (\ref{barnumber}) 
decreases as \(T^{\frac{1}{2}}\). Also the rate of 
the sphaleron transitions decreases. Below the temperature 
\(T^{*}\approx v(T^{*})\approx 100 \mbox{ GeV}\) the baryon-number is frozen 
out \cite{rubakov}. From (\ref{coefficients}), (\ref{energydif}),
and (\ref{barnumber}) we obtain for the resulting baryon-number
\be
\left.\frac{\Delta_B}{n_{\gamma}}\right|_{\mbox{now}}= 
\left( 7\;\delta_{CP}^{1}
+16\;\delta_{CP}^{2}\right)\times 10^{-5} 
\left(\frac{100 \mbox{ GeV}}{M}\right)^4,
\label{barnumbernow}\ee
where we have used \(c=1\). Also we have included the factor \(0.037\) 
to account for the 
changes in the ratio due to changes in the number of relativistic
 particle species when the universe was cooling down. 
In the standard model the magnitude of \(\delta_{CP}^{1}\), \(\delta_{CP}^{2}\)
is too small (about \(10^{-20}\)) to explain the observed matter anti-matter
asymmetry (\ref{observationratio}).
However, for extensions of the standard model 
\(\delta_{CP}^{1}, \delta_{CP}^{2}\) can be 
as large as \(10^{-3}\) and we see that (\ref{barnumbernow}) may explain  
the observed baryon-number excess (\ref{observationratio}),
without introducing a (strong) first-order phase-transition. 

\subsubsection*{Acknowledgements}
I would like to thank Jan Smit and Chris van Weert for useful discussions.

\end{document}